\documentclass[aps,prl,twocolumn,superscriptaddress,showpacs,,longbibliography]{revtex4-2}
\usepackage{amsmath,amssymb,amsfonts,bbm,float,graphics,epsfig,epstopdf,color,verbatim,tabularx,bm,multirow,hyperref,ulem,times,subfigure}
\hypersetup{
  colorlinks, linkcolor={blue},
  citecolor={blue}, urlcolor={blue}
}

\begin{document}
\title{Identifying two-dimensional topological phase transition by entanglement spectrum : A fermion Monte Carlo study}

\author{Weilun Jiang}
\email{wljiang@sxu.edu.cn}
\affiliation{State Key Laboratory of Quantum Optics Technologies and Devices, Institute of Opto-Electronics, Shanxi University, Taiyuan, 030006, China }
\affiliation{Collaborative Innovation Center of Extreme Optics, Shanxi University, Taiyuan, 030006, China }

\author{Xiaofan Luo}
\affiliation{State Key Laboratory of Quantum Optics Technologies and Devices, Institute of Opto-Electronics, Shanxi University, Taiyuan, 030006, China }

\author{Bin-Bin Mao}
\affiliation{School of Foundational Education, University of Health and Rehabilitation Sciences, Qingdao 266000, China}
\affiliation{Department of Physics, School of Science and Research Center for Industries of the Future, Westlake University, Hangzhou 310030,  China}

\author{Zheng Yan}
\email{zhengyan@westlake.edu.cn}
\affiliation{Department of Physics, School of Science and Research Center for Industries of the Future, Westlake University, Hangzhou 310030,  China}
\affiliation{Institute of Natural Sciences, Westlake Institute for Advanced Study, Hangzhou 310024, China}

\begin{abstract}
Among many types of quantum entanglement properties, the entanglement spectrum provides more abundant information than other observables. Exact diagonalization and density matrix renormalization group method could handle the system in one-dimension properly, while in higher dimension, it exceeds the capacity of the algorithms. To expand the ability of existing numerical methods, we takes a different approach via quantum Monte Carlo algorithm. By exploiting particle number and spin symmetry, we realize an efficient algorithms to solve the entanglement spectrum in the interacting fermionic system. Taking two-dimensional interacting Su-Schrieffer-Heeger as example, we verify the existence of topological phase transition under different types of many-body interactions. The calculated particle number distribution and wave-function of entanglement Hamiltonian indicate that the two belong distinct types of topological phase transitions.
\end{abstract}

\date{\today}
\maketitle

\section{Introduction} 
\label{sec:seci}
In the field of many-body calculations, phase transitions were typically described by specifying the corresponding order parameter and determining the location of the phase transition based on its value changes or scaling behavior \cite{Sachdev2011Quantum}. In recent years, with the continuous development of the field of quantum information, the concept of quantum entanglement has also been introduced into the characterization of many-body problems \cite{Horodecki2009Quantum,Amico2008Entanglement}. Entanglement describes the non-local quantum correlations within a system, and such behavior has been found to be powerful for distinguishing different phases and phase transitions \cite{Eisert2010Area,Laflorencie2016}. As a result, the study of entanglement provides a unique perspective and has become a widely concerned physical observable in current research. Here, we focus on the simplest and most studied case -- bipartite entanglement. Among various bipartite entanglement observables, the entanglement spectrum (ES) can encompass more physical information, since it serves as a spectral structure \cite{Li2008Entanglement}. A lot of previous studies have used the ES to investigate problems such as topological phases with bulk-edge correspondence \cite{Pollmann2010Entanglement,Fidkowski2010Entanglement,Prodan2010Entanglement,Yao2010Entanglement,Qi2012General,Hsieh2014Bulk,Gong2018Topological,Yu2024Universal} in different systems: spin lattices \cite{Alba2012Boundary,De2012Entanglement,Lundgren2016Universal,Calabrese2008Entanglement,Serbyn2016Power,Alba2013Entanglement,yan2023unlocking,liu2024demonstrating,wang2024sudden} and fermionic systems \cite{Fries2019Entanglement,Chung2001Density,Thomale2010Nonlocal,Herviou2019Entanglement,Eldad2022Entanglement,assaad2015stable,assaad2014entanglement,parisen2018entanglement}. These studies further offer a deeper understanding to the inherent symmetries of the model and the emergent symmetry at phase transition points \cite{Cho2017Universal,Mao2023Sampling}. It indicates that the property of ES is an extremely important observable for exploring phase of matter.

However, the study of entanglement requires partial tracing of the whole density matrix, named reduced density matrix (RDM) \cite{Ingo2003Calculation,Peschel2009Reduced}, which is not an easy task in actual many-body calculations due to the huge degree of freedom. Specifically, the existing methods for studying the ES are mainly based on density matrix renormalization group (DMRG) and exact diagonalization method (ED), or few analytical derivation methods. All these approaches become particularly difficult to be extended to the dimensions higher than one, due to the inherent algorithmic structure. Even if applied to the higher dimension, the calculation on the large system sizes are limited, because of the complexity in the full trace operation of environment. Thus the information uncovered by ES suffers from significant finite-size effects, especially, when dealing with long-range problems. Therefore, research on ES is mainly confined to one-dimensional systems, and the development of methods for studying ES in high-dimensional systems, as well as the exploration of its physical mechanisms, is still ongoing. Fortunately, the quantum Monte Carlo algorithm (QMC) is an unbiased method and has advantages in solving high-dimensional problems. In recent years, there have been several advancements in QMC algorithms for studying bipartite entanglement with high efficiency, such as entanglement entropy and R\'enyi negativity \cite{D2024Universal,Wang2023Entanglement,Mao2023Sampling,Wang2024Probably,Liu2023Non,Wu2020Entanglement,Liao2023Controllable,Wang2024Entanglement,ding2024tracking}. The advantage of these newly-proposed algorithms is that they are in principle able to handle all two or higher dimensional interacting systems between spins or fermions by assigning segment calculation. Hence, the blanks in the ES calculation by QMC are needed to be filled in.

In this work, we use the determinantal quantum Monte Carlo (DQMC) method to study the low-lying levels of ES in the interacting fermion models as the first time attempt. To optimize the computational method, we analyze and identify all possible conserved quantities to simplify the computation of the ES. We are interested in the topological problems and take the topology properties of the non-interacting Su-Schrieffer-Heeger (SSH) model as a starting point. Next, we calculate the topological phase transition of the system under the many-body Hubbard interaction. Remarkably, we implement this method in two-dimensional systems and verify its correctness, based on the conclusions from ED methods in this one-dimensional SSH-Hubbard model. We also apply the algorithm into the case of nearest-neighbor density interaction, and find the numerical results are consistent with entanglement analysis based on wave-function.

\section{Calculation of entanglement spectrum} 
We provide the details on how to treat the RDM as an feasible observables. Pioneered by Ref.\cite{Grover2013Entanglement}, within DQMC regime $\rho_M$ is expressed as the averaged RDM among configurations,
\begin{equation}
	\rho_M = \sum_s P_s \rho_{M,s} 
\end{equation}
where $s$ labels the auxiliary field, $P_s$ is the probability for each auxiliary field. Since the single auxiliary field configuration serves as the non-interacting fermionic system, $\rho_s$ now directly depends on the correlation matrix $G_s$, 
\begin{equation}
	\rho_{M, s}=C_{s, M} e^{-\mathbf{c}^{\dagger} \log \left(G_{s, M}^{-1}-\mathbbm{1}\right) \mathbf{c}} = C_{s, M} e^{-\mathbf{c}^{\dagger} H_{s,M} \mathbf{c}},
\end{equation}
$\mathbf{c} = (c_1, c_2, \cdots)$ denotes the fermionic vector, and $G_{s,M} = \langle c_i^\dagger c_j \rangle,\quad i,j \in M$ is the correlation matrix defined on $M$. $H_{s,M}$ is known as the reduced Hamiltonian or entangled Hamiltonian (EH) for configuration $s$\cite{Ingo2003Calculation}. $C_{s, M} = \text{Det} ( \mathbbm{1} - G_{s,M})$ normalize the trace of RDM to the unity. 
In QMC simulation, $\rho_{M,s}$ serves as an observable, which should be measured for as many times as possible to control the errorbar. However, the computational cost for measuring such observables is heavy, resulted in the inability to the large entangled size. To be specific, the most costly part, which comes from the exponential computation with $d_M \times d_M$ matrix diagonalization, is performed on each sampling, where $d_M = d^{N_M}$ labels the dimension of Hilbert space on region $M$. 

To simplify the complexity, we first explore the properties of the $H_{s,M}$. Considering the general interacting fermionic Hamiltonian $\mathcal{H} = \mathcal{T} + \mathcal{V}$ in DQMC, including the hopping term $\mathcal{T}$ and many-body interactions $\mathcal{V}$. The equal-time Green function matrix is expressed as\cite{Assaad2008World-line},
\begin{equation}
  G_s =\left(\mathbbm{1}+ \prod_{l = 0}^{\beta} e^{-\Delta \tau T} e^{-\Delta \tau V(l)} \right)^{-1},
\end{equation}
where the matrix $T$ and $V$ are the hopping and interaction matrix after Hubbard-Stratonovich transformation, and $\beta$ is the inversed temperature. It is noticed that $V$ varies along with the auxiliary field configuration, which could be different for each imaginary time slice labeled $l$. As a result, $G_s$ is not simply the Green function matrix of a certain non-interacting fermionic Hamiltonian, and the same applies to $G_{s,M}$.
Therefore, $H_{s,M}$ is expected to be a totally {\it featureless} matrix. 
Nevertheless, a special case is that $T$ and $V$ are already block-diagonalized, meanwhile, so are the associated $G_s, G_{s,M}, H_{s,M}$. Such situation suits well for the numerous models with spin degree of freedom or bilayer structure without interlayer hopping terms, where the ES allows for further simplification. 

In a more general sence, we notice that $H_{s,M}$ serves as a generic matrix with basis $\mathbf{c}$ and $\mathbf{c}^\dagger$, which has the same form with the non-interacting fermionic Hamiltonian with all-to-all hopping amplitude. We find the only existing symmetry, the particle number conservation, which survives owing to its commutation relation with the hopping term. From another perspective, since Hubbard-Stratonovich transformation essentially reduces higher symmetry of original Hamiltonian to U(1), we could only make use of U(1) symmetry as the subgroup of higher symmetry, which corresponds to the particle number conservation. Considering the general EH matrix with spinless fermions, its full spectrum is block-diagonalized to $N_M+1$ sectors with respect to the conserved total particle number $n$, where $N_M$ is the size of the entangled region. Each square blocked RDM has the dimension expressed on the combinatorial number $C_{N_M}^n$.

In this work, the particle number conservation and the blocked property with spin index are both considered to simplify the calculation. In the simulation, we divide the RDM into block sectors, then calculate exponentiation and store separately. Specifically, the total Hilbert space of the entangled region is now $2^{N_s \times N_M}$, where $N_s$ is the number of spin flavors. We separate the whole RDM into $(N_M+1)^{N_s}$ small blocks labeled $\{n_1, n_2, \cdots, n_s\}$ of the particle numbers of each spin flavor, with its dimension $C_{N_M}^{n_1} \times C_{N_M}^{n_2} \times \cdots \times C_{N_M}^{n_s}$. At the final step, we do the diagonalization for each block to obtain its eigenvalue-spectrum and eigenstates of each particle number $n$ and spin index $\sigma$. These simplifications guarantee the enough system size of the entangled region to approach the nature in the thermodynamics limits.

At the end of this section, we provide a rough estimation of the algorithm complexity to compare with ED algorithm. The most costly process among both algorithms is the diagonalization process to obtain eigenvalues. However, we emphasize that above blocked diagonalization properties also exists in original ED methods. The dominating simplification is the combination of the Monte Carlo algorithm. Reminiscent of the ED algorithm, the complexity is equivalent to the exponential function of the total Hilbert space, O($d^N$). Thus, the complexity of Monte Carlo algorithm is the exponential function of the entangled region size, O($d^{N_M}$). More accurately, in addition to the Monte Carlo sampling process, the ES calculation by DQMC should be O($ N^3 d^{N_M}$), which reduces the complexity from the exponential order to the algebraic order of the total system size. Therefore, we conclude that the tracing out the large degree of freedom from the environment is more applicable to solve by our optimized method.  

\section{Two-dimensional SSH model} 
\begin{figure}[h]
  \centering
  \includegraphics[width=\columnwidth]{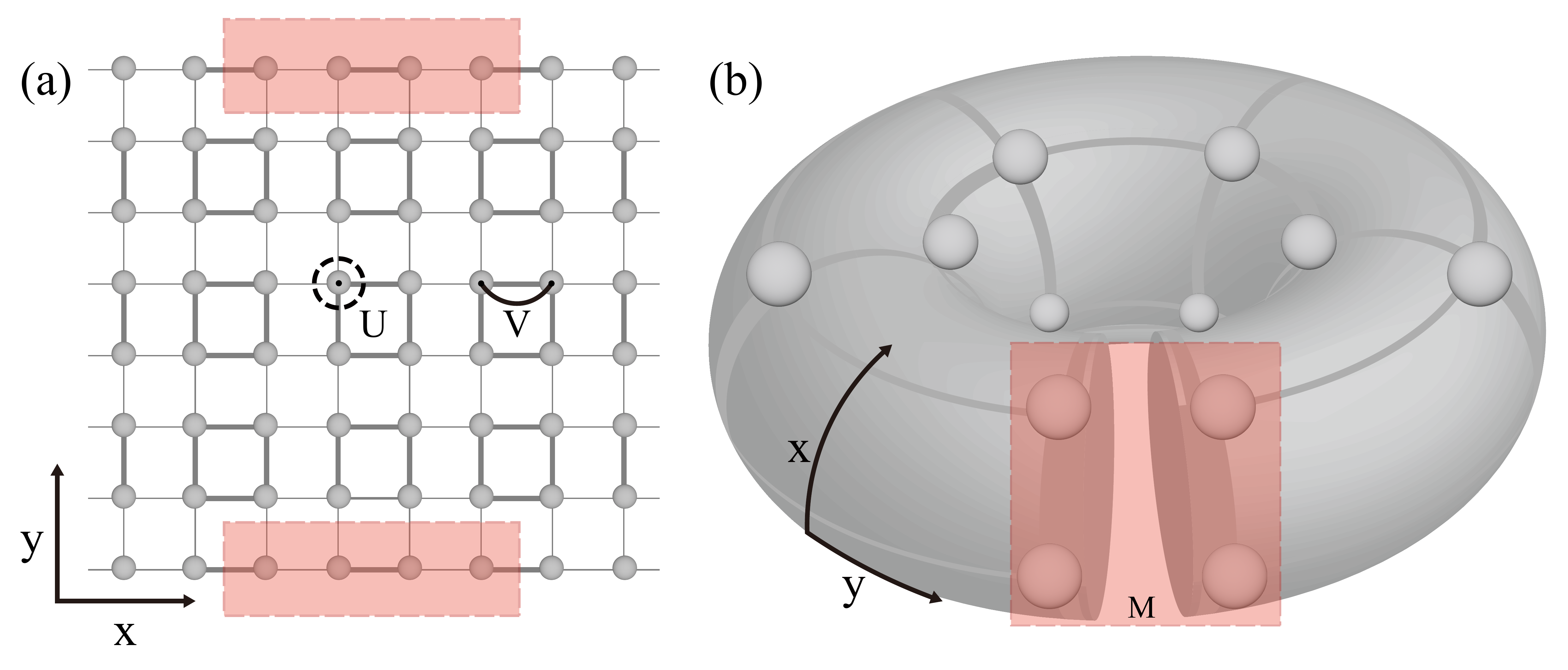}
  \caption{Sketch map of the two-dimensional SSH model. (a) The thick and thin bonds represent the weak hopping strength $t_{xa}, t_{ya}$ and strong hopping strength $t_{xb}, t_{yb}$, respectively. We choose periodic boundary conditions (PBC) along $x$-axis and open boundary conditions (OBC) along $y$-axis. The entangled region $M$ is colored by red, as an example of size $4 \times 2$. (b) The lattice sketch map in three-dimensional perspective. The OBC is set in the region $M$. }
  \label{fig1}
\end{figure}

We apply the algorithm to the entanglement study on the SSH model. The one-dimensional SSH model exhibits the topological phase transition from trivial gapped state to the topological phase with zero energy modes by tuning the ratio of two staggered coupling constants. Here, we extend the one-dimensional SSH model to two-dimension, which links several one-dimensional SSH chains with staggered hopping strength along the vertical direction, i.e. SSH-type chains along $y$-axis. On one hand, the computation on this two dimensional model enables us to reveal the superiority of our method compared with ED and DMRG methods. On the other hand, a total of four edge states contribute to the ground state degeneracy, bringing more complexity. The non-interacting Hamiltonian $H_0$ writes as
\begin{equation}
  \begin{aligned}
  H_0 &= H_x + H_y \\
  H_x &= - t_{xa} \sum_{j,i = \text{even}}^{L_x} c^{\dagger}_{(i,j)} c_{(i+1,j)}  - t_{xb} \sum_{j,i = \text{odd} }^{L_x} c^{\dagger}_{(i,j)} c_{(i+1,j) } + h.c. \\
  H_y &= - t_{ya} \sum_{i,j = \text{even}}^{L_y} c^{\dagger}_{(i,j)} c_{(i,j+1)}  - t_{yb} \sum_{i,j = \text{odd} }^{L_y} c^{\dagger}_{(i,j)} c_{(i,j+1) } + h.c. 
  \end{aligned}
\end{equation}
The sketch map of the two-dimensional SSH model is shown in Fig.\ref{fig1}. The alternating hopping strength are along both $x$ and $y$ directions, denoted as $t_{xa}, t_{xb}, t_{ya}, t_{yb}$, respectively. 

Thereafter, considering the many-body interactions, we choose two sorts of them: 1) onsite Hubbard interaction $H_U$; 2) spinless nearest-neighbor interaction $H_V$:
\begin{equation}
  \begin{aligned}
      H_{U} &= U \sum_{I} ( n_{I,\uparrow} - \frac{1}{2})( n_{I,\downarrow} - \frac{1}{2}) \\
      H_{V} &= V \sum_{\langle IJ \rangle} ( n_I - \frac{1}{2})( n_J - \frac{1}{2}), 
  \end{aligned}
\end{equation}
both of which are free of sign problem in Monte Carlo simulation. In addition, to avoid the sign-problem, we only focus on the half-filled case. The inverse temperature is chose $\beta = 100$, which is low enough to reach the ground state.  The two-dimensional square lattice site is represents by capital Roman $I, J$, with its coordinate denoted as $(i,j)$.

\section{Entanglement spectrum under the Hubbard interaction} 
We first provide the benchmark of the ES algorithm by utilizing the Hubbard interaction. As a result, the spin's degree of freedom is now introduced in the original SSH model. Previously, Ref.\cite{Ye2016Entanglement} has explored one-dimensional SSH model with Hubbard interaction by ED method. Here, we generalize the model to two-dimension, where the calculation is beyond the capacity of ED method. We hence employ DQMC method based on the interacting fermionic system, together with the optimized algorithm of ES to numerically explore its topological properties. 

\begin{figure}[h]
  \centering
  \includegraphics[width=\columnwidth]{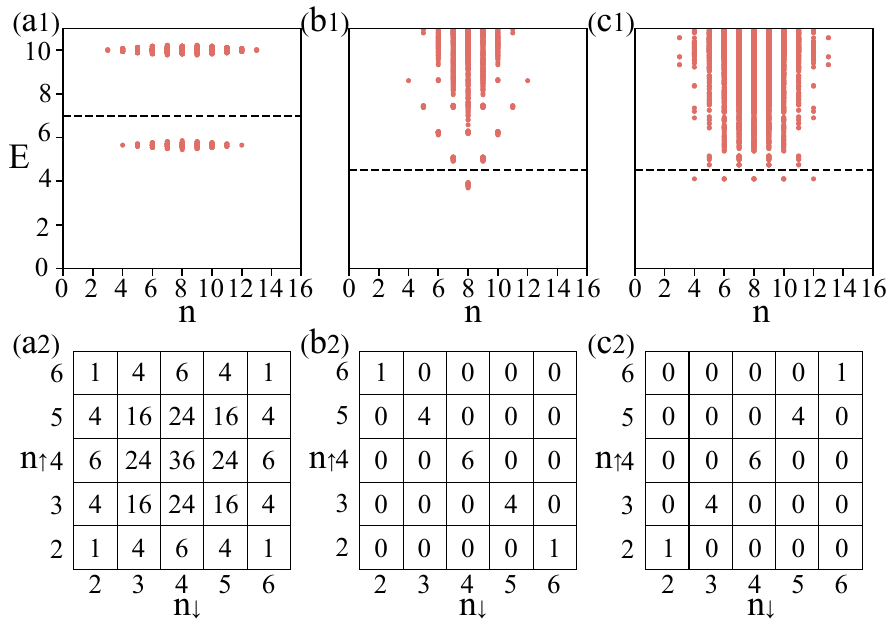}
  \caption{ES versus total particle number and eigenstates distribution versus each spin at the existence of $H_U$. From left to right panels, $U = 0, 2.0, -2.0$ respectively. The dashed line distinguishes the ground state and the first excited states. The topological phase transition occurs by adding the Hubbard interaction, manifesting ground state degeneracy changes from $256$ to $16$. We also observe the eigenstates distribution of positive and negative $U$ is distinct. All the results are calculated on a square lattice with PBC along x-axis and OBC along y-axis, as shown in Fig.\ref{fig1} (b).}
  \label{fig2}
\end{figure}

At beginning, we provide a brief analysis and the prediction for the two-dimensional SSH model with Hubbard interaction. In absence of $H_U$, the ES of the subsystem has total eight stable zeros energy modes (two edge states along both $x$ and $y$ direction, two spin's degree of freedoms), on the condition that the region is made up by cutting off two bonds on the boundaries with stronger hopping amplitude. The associated ES of non-interacting case is shown in Fig.\ref{fig2} (a1) and (a2) with 256-fold ($2^8$) ground state degeneracy. To guarantee that we are in the topological phase, we fix $t_{xa} = 0.2, t_{xb} = 0.8, t_{ya} = 0.2, t_{yb} = 0.8$, and we choose the entanglement region $M$ with size $M_x = 4$ and $M_y = 2$. Note four lattice sites are the smallest size along $x$ direction to guarantee the presence of zero energy modes by cutting the region. The total system size $L_x = 16, L_y = 8$ is large enough to avoid finite size effect generated by the total system.

After adding $H_U$, for a one-dimensional SSH model, the ground state degeneracy changes from $16$-fold (two edge modes and two spin flavors) to $4$-fold \cite{Ye2016Entanglement}. Besides, the positive and negative $U$ exhibits different distribution of the ground state spectrum with respect to the particle number. Based on the conclusion of the one-dimensional model, we expect spontaneously that in two-dimensional SSH model is treated as simple additivity of one-dimensional features. Specifically, the ground state degeneracy change form $16^2$ to $4^2$. Meanwhile, the particle number distribution for positive $U$ and negative $U$ is supposed to remain different. For example, with negative $U$, the pair of fermions with spin up and spin down tend to locate on the same site. Therefore, the pair will do or not exist on sites on the boundary together, which leads to $4$-fold topological degeneracy and equal number of spin up and spin down particles of the eigenstates \cite{Ye2016Entanglement}. It finally gives rise to the degeneracy with different total particle numbers sector, in contrast to the positive $U$ case with single particle occupied on each edge state. 

We display the QMC results in Fig.\ref{fig2} (b1) and (c1). The ES are calculated in each total particle number sector $n = n_{\uparrow} + n_{\downarrow}$  with $U = 2.0, -2.0$ respectively. At the presence of the interaction, we clearly find the ground states have 16-fold degeneracy and own a clear ES gap between the first excited state. However, on account of the finite size effect of $M_x$, the eigenvalues are not exactly equal to the putative degenerated energy levels, as shown both in the ground state and the first excited state. In any case, we conclude the ES in the thermodynamics limit has ground state topological degeneracy in both negative and positive $U$. Besides, the degeneracy distributes in different form with respect to the particle number of two $U$s as in Fig.\ref{fig2} (b2) and (c2). It is interpreted as the two copies for the particle number distribution of one-dimensional system. 

In addition, we find the high-energy part of ES behaves like a continuous spectrum without distinguishable energy gap which is totally different from the free fermion case. Physically, it means the quasiparticles have interactions in high energy part which is a typical effect of strongly correlated electrons. For the super-high part, another possible reason is the numerical error of sampling. For the former, the eigenvalues of super-high-energy part are exponentially small, which is influenced by the numerical error to a much greater extent than the lower part. Therefore, for ES calculated by QMC simulation, only the low-energy part could be characteristic and credible. 

For a normal QMC algorithm by directly simulating the system with open boundary SSH model, it is hard to recognize the ground state degeneracy, and hence distinguish the topological phase transition. Meanwhile, our ES method can provide detailed information of the wavefunctions and its associate degeneracy through bulk-edge correspondence.

\section{Entanglement spectrum under the density interaction} %
\begin{figure}[h]
  \centering
  \includegraphics[width=0.8\columnwidth]{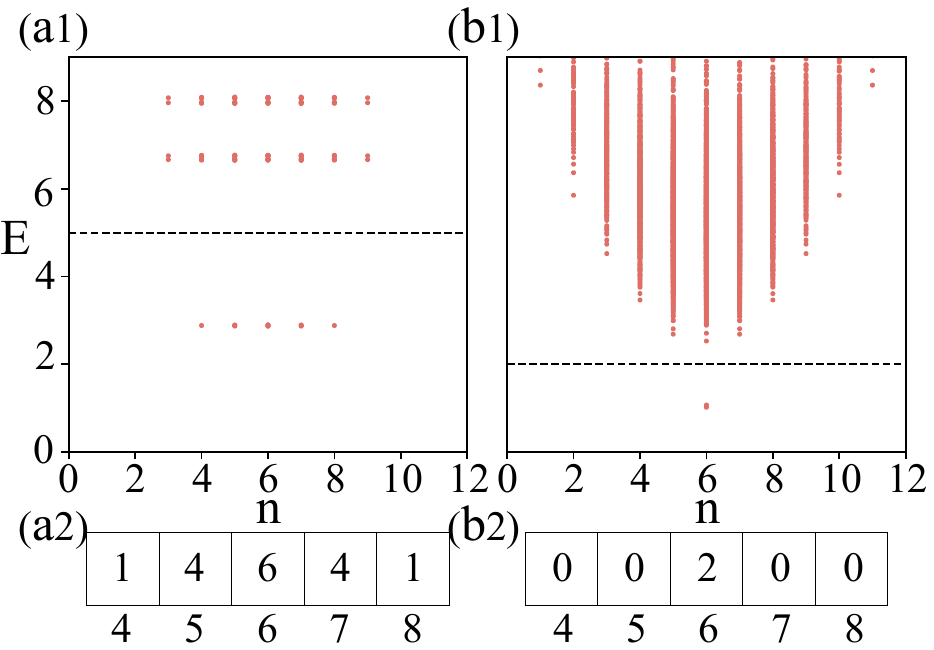}
  \caption{ES versus total particle number and eigenstates distribution versus each spin at the existence of $H_V$. $V = 0, 1$ for the left and right panels, respectively. The red line distinguishes the ground state and the first excited states. The topological phase transition occurs by adding the $H_V$ interaction, manifesting ground state degeneracy changes from $16$ to $2$.}
  \label{fig3}
\end{figure}

As an application of this optimized algorithm, we explore the topological phase transition by applying spinless density interaction, termed as $H_V$ in this section. Different from the Hubbard-type interaction, $H_V$ term directly couples the edge modes by its nearest-neighbor density interactions in two-dimensional case, as shown in the sketch map of Fig.\ref{fig1} (a). Since we consider the spinless system, we could reach larger size of $M$ as $M_x = 6, M_y = 2$, and the non-interacting ground state degeneracy is now $16$-fold as shown in Fig.\ref{fig3} (a1) and (a2). Beyond that, to avoid the sign problem for such interaction, we only focus $V > 0$ region and choose $V = 1$ for the computation. Fig. \ref{fig3} (b1) and (b2) show the QMC results. We numerically observe the ground state degeneracy reduces from $16$ to $2$, entirely different from the positive Hubbard interaction, and the two degenerated ground states are both half-filled. 

\begin{figure}[h]
  \centering
  \includegraphics[width=\columnwidth]{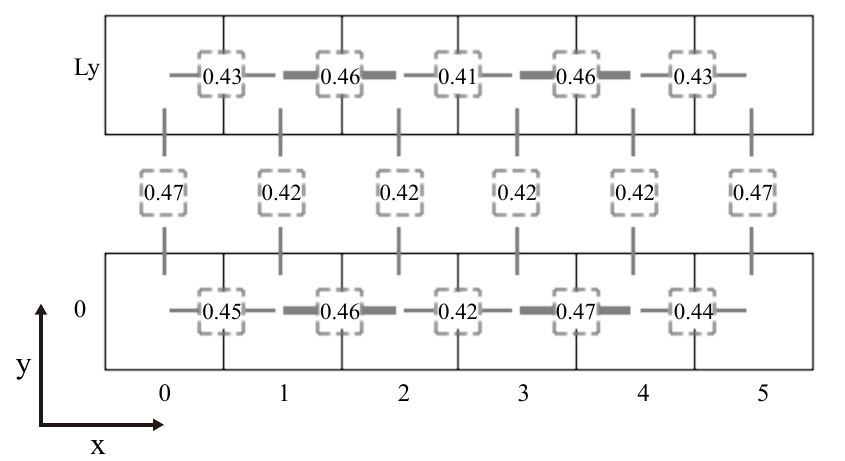}
  \caption{Site-resolved CDW correlation function for entangled region. The thin and thick bonds represent the weak and strong hopping strength, respectively. All average values are closed to 0.5, dated from strong global CDW order. Furthermore, we find $d_{(1,0)(2,0)}$ and $d_{(3,0)(4,0)}$ is larger than $d_{(0,0)(1,0)}$, $d_{(2,0)(3,0)}$ and $d_{(4,0)(5,0)}$, consistent with the alternating hopping strength of original Hamiltonian. Note both two entangled wavefuntions are indistinguishable from the studied observables.}
  \label{fig4}
\end{figure}

The repulsive density interaction promotes the charge density wave (CDW) order for spinless fermion. However, in the $V=0$ limit, the degeneracy stretches across various sectors of total number. When adding the interaction, we expect only the half-filled degeneracies survive. The degeneracy in one-dimensional case is expected to transfer from $4$ to $2$. Moreover, in two-dimension, the CDW order forms the staggered configurations of the particle number along both $x$ and $y$ direction, thus only two-fold degeneracy exists. To further verify the correctness, it is available to extract the ground state wavefunction of the EH.
The results are expressed in Fig.\ref{fig4}. We find the average value of density on each site $n_i$ is 0.5. Then we measure the average value of the site-resolved nearest-neighbor CDW density correlation $d_{IJ} = n_I ( 1 - n_J)$ of the two eigenstates. At the presence of $H_V$, $d_{(i,j) (i+1,j)}$ is closed to 0.5 for both degenerated state, indicating the strong CDW order along $x$ direction for the ground state of EH. Furthermore, the density correlation between two edge modes $d_{(i,1) (i,L_y)}$ is also approximate to 0.5, even though they are not in direct connection due to the OBC along $y$-axis, which reveals a global CDW order along both direction. We hence expect the two degenerate states are the linear combination of $| 0,1,0,1,\cdots \rangle$ and $| 1,0,1,0,\cdots \rangle$ in the occupying number representation, which further demonstrates the degenerace of the ES here. In conclusion, we identify the topological phase transition of two-dimensional SSH model from $16$-fold to $2$-fold by adding the density interaction. An interesting thing is that the phase transition in the ground-state of EH also reflects the transition in the ground-state of original Hamiltonian, which answers a long-standing controversy in the field~\cite{PhysRevLett.113.060501,wang2024sudden}

\section{Summary and outlook} 
We report an efficient way to numerically calculate the ES and its entangled wavefunction of two dimensional interacting system based on the large-scale QMC algorithms. We analyze its algorithm optimization, and leverage particle number and spin conservation. Then we verify the algorithm with two-dimensional SSH model with Hubbard interaction. In addition, we investigate the topological phase transition of SSH model under density-density interaction, and offer logical explanations to this. Importantly, our scheme could be extended to two or higher dimension with large system size, which surpasses the computation capacity of other exact or numerical method. However, the current available size of entangled region is limited to small subsystem. To reach larger entangled region, one could use some approximation technique to extract the ground state or low excited state information. Nonetheless, we show the ability for QMC algorithm to study ES, which gives researchers more avenues in numerical techniques. Our method thus provides an opportunity to explore the physical mechanism of interacting fermion system via entanglement properties. 

\section*{Acknowledgment}
This work is supported by National Natural Science Foundation of China (Project No. 12404275), and the fundamental research program of Shanxi province (Project No. 202403021212015).  ZY acknowledges the start-up funding of Westlake University. The authors thank the high-performance computing center of Westlake University and Beijng PARATERA Tech Co.,Ltd. for providing HPC resources.

\bibliography{main}

\begin{thebibliography}{47}%
\makeatletter
\providecommand \@ifxundefined [1]{%
 \@ifx{#1\undefined}
}%
\providecommand \@ifnum [1]{%
 \ifnum #1\expandafter \@firstoftwo
 \else \expandafter \@secondoftwo
 \fi
}%
\providecommand \@ifx [1]{%
 \ifx #1\expandafter \@firstoftwo
 \else \expandafter \@secondoftwo
 \fi
}%
\providecommand \natexlab [1]{#1}%
\providecommand \enquote  [1]{``#1''}%
\providecommand \bibnamefont  [1]{#1}%
\providecommand \bibfnamefont [1]{#1}%
\providecommand \citenamefont [1]{#1}%
\providecommand \href@noop [0]{\@secondoftwo}%
\providecommand \href [0]{\begingroup \@sanitize@url \@href}%
\providecommand \@href[1]{\@@startlink{#1}\@@href}%
\providecommand \@@href[1]{\endgroup#1\@@endlink}%
\providecommand \@sanitize@url [0]{\catcode `\\12\catcode `\$12\catcode `\&12\catcode `\#12\catcode `\^12\catcode `\_12\catcode `\%12\relax}%
\providecommand \@@startlink[1]{}%
\providecommand \@@endlink[0]{}%
\providecommand \url  [0]{\begingroup\@sanitize@url \@url }%
\providecommand \@url [1]{\endgroup\@href {#1}{\urlprefix }}%
\providecommand \urlprefix  [0]{URL }%
\providecommand \Eprint [0]{\href }%
\providecommand \doibase [0]{https://doi.org/}%
\providecommand \selectlanguage [0]{\@gobble}%
\providecommand \bibinfo  [0]{\@secondoftwo}%
\providecommand \bibfield  [0]{\@secondoftwo}%
\providecommand \translation [1]{[#1]}%
\providecommand \BibitemOpen [0]{}%
\providecommand \bibitemStop [0]{}%
\providecommand \bibitemNoStop [0]{.\EOS\space}%
\providecommand \EOS [0]{\spacefactor3000\relax}%
\providecommand \BibitemShut  [1]{\csname bibitem#1\endcsname}%
\let\auto@bib@innerbib\@empty
\bibitem [{\citenamefont {Sachdev}(2011)}]{Sachdev2011Quantum}%
  \BibitemOpen
  \bibfield  {author} {\bibinfo {author} {\bibfnamefont {S.}~\bibnamefont {Sachdev}},\ }\href {https://doi.org/10.1017/CBO9780511973765} {\emph {\bibinfo {title} {Quantum Phase Transitions}}},\ \bibinfo {edition} {2nd}\ ed.\ (\bibinfo  {publisher} {Cambridge University Press},\ \bibinfo {year} {2011})\BibitemShut {NoStop}%
\bibitem [{\citenamefont {Horodecki}\ \emph {et~al.}(2009)\citenamefont {Horodecki}, \citenamefont {Horodecki}, \citenamefont {Horodecki},\ and\ \citenamefont {Horodecki}}]{Horodecki2009Quantum}%
  \BibitemOpen
  \bibfield  {author} {\bibinfo {author} {\bibfnamefont {R.}~\bibnamefont {Horodecki}}, \bibinfo {author} {\bibfnamefont {P.}~\bibnamefont {Horodecki}}, \bibinfo {author} {\bibfnamefont {M.}~\bibnamefont {Horodecki}},\ and\ \bibinfo {author} {\bibfnamefont {K.}~\bibnamefont {Horodecki}},\ }\bibfield  {title} {\bibinfo {title} {Quantum entanglement},\ }\href {https://doi.org/10.1103/RevModPhys.81.865} {\bibfield  {journal} {\bibinfo  {journal} {Reviews of Modern Physics}\ }\textbf {\bibinfo {volume} {81}},\ \bibinfo {pages} {865} (\bibinfo {year} {2009})}\BibitemShut {NoStop}%
\bibitem [{\citenamefont {Amico}\ \emph {et~al.}(2008)\citenamefont {Amico}, \citenamefont {Fazio}, \citenamefont {Osterloh},\ and\ \citenamefont {Vedral}}]{Amico2008Entanglement}%
  \BibitemOpen
  \bibfield  {author} {\bibinfo {author} {\bibfnamefont {L.}~\bibnamefont {Amico}}, \bibinfo {author} {\bibfnamefont {R.}~\bibnamefont {Fazio}}, \bibinfo {author} {\bibfnamefont {A.}~\bibnamefont {Osterloh}},\ and\ \bibinfo {author} {\bibfnamefont {V.}~\bibnamefont {Vedral}},\ }\bibfield  {title} {\bibinfo {title} {Entanglement in many-body systems},\ }\href {https://doi.org/10.1103/RevModPhys.80.517} {\bibfield  {journal} {\bibinfo  {journal} {Reviews of Modern Physics}\ }\textbf {\bibinfo {volume} {80}},\ \bibinfo {pages} {517} (\bibinfo {year} {2008})}\BibitemShut {NoStop}%
\bibitem [{\citenamefont {Eisert}\ \emph {et~al.}(2010)\citenamefont {Eisert}, \citenamefont {Cramer},\ and\ \citenamefont {Plenio}}]{Eisert2010Area}%
  \BibitemOpen
  \bibfield  {author} {\bibinfo {author} {\bibfnamefont {J.}~\bibnamefont {Eisert}}, \bibinfo {author} {\bibfnamefont {M.}~\bibnamefont {Cramer}},\ and\ \bibinfo {author} {\bibfnamefont {M.~B.}\ \bibnamefont {Plenio}},\ }\bibfield  {title} {\bibinfo {title} {Colloquium: Area laws for the entanglement entropy},\ }\href {https://doi.org/10.1103/RevModPhys.82.277} {\bibfield  {journal} {\bibinfo  {journal} {Reviews of Modern Physics}\ }\textbf {\bibinfo {volume} {82}},\ \bibinfo {pages} {277} (\bibinfo {year} {2010})}\BibitemShut {NoStop}%
\bibitem [{\citenamefont {Laflorencie}(2016)}]{Laflorencie2016}%
  \BibitemOpen
  \bibfield  {author} {\bibinfo {author} {\bibfnamefont {N.}~\bibnamefont {Laflorencie}},\ }\bibfield  {title} {\bibinfo {title} {Quantum entanglement in condensed matter systems},\ }\href {https://doi.org/https://doi.org/10.1016/j.physrep.2016.06.008} {\bibfield  {journal} {\bibinfo  {journal} {Physics Reports}\ }\textbf {\bibinfo {volume} {646}},\ \bibinfo {pages} {1} (\bibinfo {year} {2016})},\ \bibinfo {note} {quantum entanglement in condensed matter systems}\BibitemShut {NoStop}%
\bibitem [{\citenamefont {Li}\ and\ \citenamefont {Haldane}(2008)}]{Li2008Entanglement}%
  \BibitemOpen
  \bibfield  {author} {\bibinfo {author} {\bibfnamefont {H.}~\bibnamefont {Li}}\ and\ \bibinfo {author} {\bibfnamefont {F.~D.~M.}\ \bibnamefont {Haldane}},\ }\bibfield  {title} {\bibinfo {title} {Entanglement spectrum as a generalization of entanglement entropy: Identification of topological order in non-abelian fractional quantum hall effect states},\ }\bibfield  {journal} {\bibinfo  {journal} {Physical Review Letters}\ }\textbf {\bibinfo {volume} {101}},\ \href {https://doi.org/10.1103/PhysRevLett.101.010504} {10.1103/PhysRevLett.101.010504} (\bibinfo {year} {2008})\BibitemShut {NoStop}%
\bibitem [{\citenamefont {Pollmann}\ \emph {et~al.}(2010)\citenamefont {Pollmann}, \citenamefont {Turner}, \citenamefont {Berg},\ and\ \citenamefont {Oshikawa}}]{Pollmann2010Entanglement}%
  \BibitemOpen
  \bibfield  {author} {\bibinfo {author} {\bibfnamefont {F.}~\bibnamefont {Pollmann}}, \bibinfo {author} {\bibfnamefont {A.~M.}\ \bibnamefont {Turner}}, \bibinfo {author} {\bibfnamefont {E.}~\bibnamefont {Berg}},\ and\ \bibinfo {author} {\bibfnamefont {M.}~\bibnamefont {Oshikawa}},\ }\bibfield  {title} {\bibinfo {title} {Entanglement spectrum of a topological phase in one dimension},\ }\bibfield  {journal} {\bibinfo  {journal} {Physical Review B}\ }\textbf {\bibinfo {volume} {81}},\ \href {https://doi.org/10.1103/PhysRevB.81.064439} {10.1103/PhysRevB.81.064439} (\bibinfo {year} {2010})\BibitemShut {NoStop}%
\bibitem [{\citenamefont {Fidkowski}(2010)}]{Fidkowski2010Entanglement}%
  \BibitemOpen
  \bibfield  {author} {\bibinfo {author} {\bibfnamefont {L.}~\bibnamefont {Fidkowski}},\ }\bibfield  {title} {\bibinfo {title} {Entanglement spectrum of topological insulators and superconductors},\ }\href {https://doi.org/10.1103/PhysRevLett.104.130502} {\bibfield  {journal} {\bibinfo  {journal} {Phys Rev Lett}\ }\textbf {\bibinfo {volume} {104}},\ \bibinfo {pages} {130502} (\bibinfo {year} {2010})}\BibitemShut {NoStop}%
\bibitem [{\citenamefont {Prodan}\ \emph {et~al.}(2010)\citenamefont {Prodan}, \citenamefont {Hughes},\ and\ \citenamefont {Bernevig}}]{Prodan2010Entanglement}%
  \BibitemOpen
  \bibfield  {author} {\bibinfo {author} {\bibfnamefont {E.}~\bibnamefont {Prodan}}, \bibinfo {author} {\bibfnamefont {T.~L.}\ \bibnamefont {Hughes}},\ and\ \bibinfo {author} {\bibfnamefont {B.~A.}\ \bibnamefont {Bernevig}},\ }\bibfield  {title} {\bibinfo {title} {Entanglement spectrum of a disordered topological chern insulator},\ }\href {https://doi.org/10.1103/PhysRevLett.105.115501} {\bibfield  {journal} {\bibinfo  {journal} {Phys Rev Lett}\ }\textbf {\bibinfo {volume} {105}},\ \bibinfo {pages} {115501} (\bibinfo {year} {2010})}\BibitemShut {NoStop}%
\bibitem [{\citenamefont {Yao}\ and\ \citenamefont {Qi}(2010)}]{Yao2010Entanglement}%
  \BibitemOpen
  \bibfield  {author} {\bibinfo {author} {\bibfnamefont {H.}~\bibnamefont {Yao}}\ and\ \bibinfo {author} {\bibfnamefont {X.~L.}\ \bibnamefont {Qi}},\ }\bibfield  {title} {\bibinfo {title} {Entanglement entropy and entanglement spectrum of the kitaev model},\ }\href {https://doi.org/10.1103/PhysRevLett.105.080501} {\bibfield  {journal} {\bibinfo  {journal} {Phys Rev Lett}\ }\textbf {\bibinfo {volume} {105}},\ \bibinfo {pages} {080501} (\bibinfo {year} {2010})}\BibitemShut {NoStop}%
\bibitem [{\citenamefont {Qi}\ \emph {et~al.}(2012)\citenamefont {Qi}, \citenamefont {Katsura},\ and\ \citenamefont {Ludwig}}]{Qi2012General}%
  \BibitemOpen
  \bibfield  {author} {\bibinfo {author} {\bibfnamefont {X.~L.}\ \bibnamefont {Qi}}, \bibinfo {author} {\bibfnamefont {H.}~\bibnamefont {Katsura}},\ and\ \bibinfo {author} {\bibfnamefont {A.~W.}\ \bibnamefont {Ludwig}},\ }\bibfield  {title} {\bibinfo {title} {General relationship between the entanglement spectrum and the edge state spectrum of topological quantum states},\ }\href {https://doi.org/10.1103/PhysRevLett.108.196402} {\bibfield  {journal} {\bibinfo  {journal} {Phys Rev Lett}\ }\textbf {\bibinfo {volume} {108}},\ \bibinfo {pages} {196402} (\bibinfo {year} {2012})}\BibitemShut {NoStop}%
\bibitem [{\citenamefont {Hsieh}\ and\ \citenamefont {Fu}(2014)}]{Hsieh2014Bulk}%
  \BibitemOpen
  \bibfield  {author} {\bibinfo {author} {\bibfnamefont {T.~H.}\ \bibnamefont {Hsieh}}\ and\ \bibinfo {author} {\bibfnamefont {L.}~\bibnamefont {Fu}},\ }\bibfield  {title} {\bibinfo {title} {Bulk entanglement spectrum reveals quantum criticality within a topological state},\ }\href {https://doi.org/10.1103/PhysRevLett.113.106801} {\bibfield  {journal} {\bibinfo  {journal} {Phys Rev Lett}\ }\textbf {\bibinfo {volume} {113}},\ \bibinfo {pages} {106801} (\bibinfo {year} {2014})}\BibitemShut {NoStop}%
\bibitem [{\citenamefont {Gong}\ and\ \citenamefont {Ueda}(2018)}]{Gong2018Topological}%
  \BibitemOpen
  \bibfield  {author} {\bibinfo {author} {\bibfnamefont {Z.}~\bibnamefont {Gong}}\ and\ \bibinfo {author} {\bibfnamefont {M.}~\bibnamefont {Ueda}},\ }\bibfield  {title} {\bibinfo {title} {Topological entanglement-spectrum crossing in quench dynamics},\ }\href {https://doi.org/10.1103/PhysRevLett.121.250601} {\bibfield  {journal} {\bibinfo  {journal} {Phys Rev Lett}\ }\textbf {\bibinfo {volume} {121}},\ \bibinfo {pages} {250601} (\bibinfo {year} {2018})}\BibitemShut {NoStop}%
\bibitem [{\citenamefont {Yu}\ \emph {et~al.}(2024)\citenamefont {Yu}, \citenamefont {Yang}, \citenamefont {Lin},\ and\ \citenamefont {Jian}}]{Yu2024Universal}%
  \BibitemOpen
  \bibfield  {author} {\bibinfo {author} {\bibfnamefont {X.-J.}\ \bibnamefont {Yu}}, \bibinfo {author} {\bibfnamefont {S.}~\bibnamefont {Yang}}, \bibinfo {author} {\bibfnamefont {H.-Q.}\ \bibnamefont {Lin}},\ and\ \bibinfo {author} {\bibfnamefont {S.-K.}\ \bibnamefont {Jian}},\ }\bibfield  {title} {\bibinfo {title} {Universal entanglement spectrum in one-dimensional gapless symmetry protected topological states},\ }\bibfield  {journal} {\bibinfo  {journal} {Physical Review Letters}\ }\textbf {\bibinfo {volume} {133}},\ \href {https://doi.org/10.1103/PhysRevLett.133.026601} {10.1103/PhysRevLett.133.026601} (\bibinfo {year} {2024})\BibitemShut {NoStop}%
\bibitem [{\citenamefont {Alba}\ \emph {et~al.}(2012)\citenamefont {Alba}, \citenamefont {Haque},\ and\ \citenamefont {Läuchli}}]{Alba2012Boundary}%
  \BibitemOpen
  \bibfield  {author} {\bibinfo {author} {\bibfnamefont {V.}~\bibnamefont {Alba}}, \bibinfo {author} {\bibfnamefont {M.}~\bibnamefont {Haque}},\ and\ \bibinfo {author} {\bibfnamefont {A.~M.}\ \bibnamefont {Läuchli}},\ }\bibfield  {title} {\bibinfo {title} {Boundary-locality and perturbative structure of entanglement spectra in gapped systems},\ }\bibfield  {journal} {\bibinfo  {journal} {Physical Review Letters}\ }\textbf {\bibinfo {volume} {108}},\ \href {https://doi.org/10.1103/PhysRevLett.108.227201} {10.1103/PhysRevLett.108.227201} (\bibinfo {year} {2012})\BibitemShut {NoStop}%
\bibitem [{\citenamefont {De~Chiara}\ \emph {et~al.}(2012)\citenamefont {De~Chiara}, \citenamefont {Lepori}, \citenamefont {Lewenstein},\ and\ \citenamefont {Sanpera}}]{De2012Entanglement}%
  \BibitemOpen
  \bibfield  {author} {\bibinfo {author} {\bibfnamefont {G.}~\bibnamefont {De~Chiara}}, \bibinfo {author} {\bibfnamefont {L.}~\bibnamefont {Lepori}}, \bibinfo {author} {\bibfnamefont {M.}~\bibnamefont {Lewenstein}},\ and\ \bibinfo {author} {\bibfnamefont {A.}~\bibnamefont {Sanpera}},\ }\bibfield  {title} {\bibinfo {title} {Entanglement spectrum, critical exponents, and order parameters in quantum spin chains},\ }\href {https://doi.org/10.1103/PhysRevLett.109.237208} {\bibfield  {journal} {\bibinfo  {journal} {Phys Rev Lett}\ }\textbf {\bibinfo {volume} {109}},\ \bibinfo {pages} {237208} (\bibinfo {year} {2012})}\BibitemShut {NoStop}%
\bibitem [{\citenamefont {Lundgren}\ \emph {et~al.}(2016)\citenamefont {Lundgren}, \citenamefont {Blair}, \citenamefont {Laurell}, \citenamefont {Regnault}, \citenamefont {Fiete}, \citenamefont {Greiter},\ and\ \citenamefont {Thomale}}]{Lundgren2016Universal}%
  \BibitemOpen
  \bibfield  {author} {\bibinfo {author} {\bibfnamefont {R.}~\bibnamefont {Lundgren}}, \bibinfo {author} {\bibfnamefont {J.}~\bibnamefont {Blair}}, \bibinfo {author} {\bibfnamefont {P.}~\bibnamefont {Laurell}}, \bibinfo {author} {\bibfnamefont {N.}~\bibnamefont {Regnault}}, \bibinfo {author} {\bibfnamefont {G.~A.}\ \bibnamefont {Fiete}}, \bibinfo {author} {\bibfnamefont {M.}~\bibnamefont {Greiter}},\ and\ \bibinfo {author} {\bibfnamefont {R.}~\bibnamefont {Thomale}},\ }\bibfield  {title} {\bibinfo {title} {Universal entanglement spectra in critical spin chains},\ }\bibfield  {journal} {\bibinfo  {journal} {Physical Review B}\ }\textbf {\bibinfo {volume} {94}},\ \href {https://doi.org/10.1103/PhysRevB.94.081112} {10.1103/PhysRevB.94.081112} (\bibinfo {year} {2016})\BibitemShut {NoStop}%
\bibitem [{\citenamefont {Calabrese}\ and\ \citenamefont {Lefevre}(2008)}]{Calabrese2008Entanglement}%
  \BibitemOpen
  \bibfield  {author} {\bibinfo {author} {\bibfnamefont {P.}~\bibnamefont {Calabrese}}\ and\ \bibinfo {author} {\bibfnamefont {A.}~\bibnamefont {Lefevre}},\ }\bibfield  {title} {\bibinfo {title} {Entanglement spectrum in one-dimensional systems},\ }\bibfield  {journal} {\bibinfo  {journal} {Physical Review A}\ }\textbf {\bibinfo {volume} {78}},\ \href {https://doi.org/10.1103/PhysRevA.78.032329} {10.1103/PhysRevA.78.032329} (\bibinfo {year} {2008})\BibitemShut {NoStop}%
\bibitem [{\citenamefont {Serbyn}\ \emph {et~al.}(2016)\citenamefont {Serbyn}, \citenamefont {Michailidis}, \citenamefont {Abanin},\ and\ \citenamefont {Papic}}]{Serbyn2016Power}%
  \BibitemOpen
  \bibfield  {author} {\bibinfo {author} {\bibfnamefont {M.}~\bibnamefont {Serbyn}}, \bibinfo {author} {\bibfnamefont {A.~A.}\ \bibnamefont {Michailidis}}, \bibinfo {author} {\bibfnamefont {D.~A.}\ \bibnamefont {Abanin}},\ and\ \bibinfo {author} {\bibfnamefont {Z.}~\bibnamefont {Papic}},\ }\bibfield  {title} {\bibinfo {title} {Power-law entanglement spectrum in many-body localized phases},\ }\href {https://doi.org/10.1103/PhysRevLett.117.160601} {\bibfield  {journal} {\bibinfo  {journal} {Phys Rev Lett}\ }\textbf {\bibinfo {volume} {117}},\ \bibinfo {pages} {160601} (\bibinfo {year} {2016})}\BibitemShut {NoStop}%
\bibitem [{\citenamefont {Alba}\ \emph {et~al.}(2013)\citenamefont {Alba}, \citenamefont {Haque},\ and\ \citenamefont {L\"auchli}}]{Alba2013Entanglement}%
  \BibitemOpen
  \bibfield  {author} {\bibinfo {author} {\bibfnamefont {V.}~\bibnamefont {Alba}}, \bibinfo {author} {\bibfnamefont {M.}~\bibnamefont {Haque}},\ and\ \bibinfo {author} {\bibfnamefont {A.~M.}\ \bibnamefont {L\"auchli}},\ }\bibfield  {title} {\bibinfo {title} {Entanglement spectrum of the two-dimensional bose-hubbard model},\ }\href {https://doi.org/10.1103/PhysRevLett.110.260403} {\bibfield  {journal} {\bibinfo  {journal} {Phys. Rev. Lett.}\ }\textbf {\bibinfo {volume} {110}},\ \bibinfo {pages} {260403} (\bibinfo {year} {2013})}\BibitemShut {NoStop}%
\bibitem [{\citenamefont {Yan}\ and\ \citenamefont {Meng}(2023)}]{yan2023unlocking}%
  \BibitemOpen
  \bibfield  {author} {\bibinfo {author} {\bibfnamefont {Z.}~\bibnamefont {Yan}}\ and\ \bibinfo {author} {\bibfnamefont {Z.~Y.}\ \bibnamefont {Meng}},\ }\bibfield  {title} {\bibinfo {title} {Unlocking the general relationship between energy and entanglement spectra via the wormhole effect},\ }\href@noop {} {\bibfield  {journal} {\bibinfo  {journal} {Nature Communications}\ }\textbf {\bibinfo {volume} {14}},\ \bibinfo {pages} {2360} (\bibinfo {year} {2023})}\BibitemShut {NoStop}%
\bibitem [{\citenamefont {Liu}\ \emph {et~al.}(2024)\citenamefont {Liu}, \citenamefont {Huang}, \citenamefont {Yan},\ and\ \citenamefont {Yao}}]{liu2024demonstrating}%
  \BibitemOpen
  \bibfield  {author} {\bibinfo {author} {\bibfnamefont {Z.}~\bibnamefont {Liu}}, \bibinfo {author} {\bibfnamefont {R.-Z.}\ \bibnamefont {Huang}}, \bibinfo {author} {\bibfnamefont {Z.}~\bibnamefont {Yan}},\ and\ \bibinfo {author} {\bibfnamefont {D.-X.}\ \bibnamefont {Yao}},\ }\bibfield  {title} {\bibinfo {title} {Demonstrating the wormhole mechanism of the entanglement spectrum via a perturbed boundary},\ }\href@noop {} {\bibfield  {journal} {\bibinfo  {journal} {Physical Review B}\ }\textbf {\bibinfo {volume} {109}},\ \bibinfo {pages} {094416} (\bibinfo {year} {2024})}\BibitemShut {NoStop}%
\bibitem [{\citenamefont {Wang}\ \emph {et~al.}(2024{\natexlab{a}})\citenamefont {Wang}, \citenamefont {Yang}, \citenamefont {Mao}, \citenamefont {Cheng},\ and\ \citenamefont {Yan}}]{wang2024sudden}%
  \BibitemOpen
  \bibfield  {author} {\bibinfo {author} {\bibfnamefont {Z.}~\bibnamefont {Wang}}, \bibinfo {author} {\bibfnamefont {S.}~\bibnamefont {Yang}}, \bibinfo {author} {\bibfnamefont {B.-B.}\ \bibnamefont {Mao}}, \bibinfo {author} {\bibfnamefont {M.}~\bibnamefont {Cheng}},\ and\ \bibinfo {author} {\bibfnamefont {Z.}~\bibnamefont {Yan}},\ }\bibfield  {title} {\bibinfo {title} {Sudden change in entanglement hamiltonian: Phase diagram of an ising entanglement hamiltonian},\ }\href@noop {} {\bibfield  {journal} {\bibinfo  {journal} {arXiv preprint arXiv:2410.10090}\ } (\bibinfo {year} {2024}{\natexlab{a}})}\BibitemShut {NoStop}%
\bibitem [{\citenamefont {Fries}\ and\ \citenamefont {Reyes}(2019)}]{Fries2019Entanglement}%
  \BibitemOpen
  \bibfield  {author} {\bibinfo {author} {\bibfnamefont {P.}~\bibnamefont {Fries}}\ and\ \bibinfo {author} {\bibfnamefont {I.~A.}\ \bibnamefont {Reyes}},\ }\bibfield  {title} {\bibinfo {title} {Entanglement spectrum of chiral fermions on the torus},\ }\href {https://doi.org/10.1103/PhysRevLett.123.211603} {\bibfield  {journal} {\bibinfo  {journal} {Phys Rev Lett}\ }\textbf {\bibinfo {volume} {123}},\ \bibinfo {pages} {211603} (\bibinfo {year} {2019})}\BibitemShut {NoStop}%
\bibitem [{\citenamefont {Chung}\ and\ \citenamefont {Peschel}(2001)}]{Chung2001Density}%
  \BibitemOpen
  \bibfield  {author} {\bibinfo {author} {\bibfnamefont {M.-C.}\ \bibnamefont {Chung}}\ and\ \bibinfo {author} {\bibfnamefont {I.}~\bibnamefont {Peschel}},\ }\bibfield  {title} {\bibinfo {title} {Density-matrix spectra of solvable fermionic systems},\ }\bibfield  {journal} {\bibinfo  {journal} {Physical Review B}\ }\textbf {\bibinfo {volume} {64}},\ \href {https://doi.org/10.1103/PhysRevB.64.064412} {10.1103/PhysRevB.64.064412} (\bibinfo {year} {2001})\BibitemShut {NoStop}%
\bibitem [{\citenamefont {Thomale}\ \emph {et~al.}(2010)\citenamefont {Thomale}, \citenamefont {Arovas},\ and\ \citenamefont {Bernevig}}]{Thomale2010Nonlocal}%
  \BibitemOpen
  \bibfield  {author} {\bibinfo {author} {\bibfnamefont {R.}~\bibnamefont {Thomale}}, \bibinfo {author} {\bibfnamefont {D.~P.}\ \bibnamefont {Arovas}},\ and\ \bibinfo {author} {\bibfnamefont {B.~A.}\ \bibnamefont {Bernevig}},\ }\bibfield  {title} {\bibinfo {title} {Nonlocal order in gapless systems: entanglement spectrum in spin chains},\ }\href {https://doi.org/10.1103/PhysRevLett.105.116805} {\bibfield  {journal} {\bibinfo  {journal} {Phys Rev Lett}\ }\textbf {\bibinfo {volume} {105}},\ \bibinfo {pages} {116805} (\bibinfo {year} {2010})}\BibitemShut {NoStop}%
\bibitem [{\citenamefont {Herviou}\ \emph {et~al.}(2019)\citenamefont {Herviou}, \citenamefont {Regnault},\ and\ \citenamefont {Bardarson}}]{Herviou2019Entanglement}%
  \BibitemOpen
  \bibfield  {author} {\bibinfo {author} {\bibfnamefont {L.}~\bibnamefont {Herviou}}, \bibinfo {author} {\bibfnamefont {N.}~\bibnamefont {Regnault}},\ and\ \bibinfo {author} {\bibfnamefont {J.~H.}\ \bibnamefont {Bardarson}},\ }\bibfield  {title} {\bibinfo {title} {Entanglement spectrum and symmetries in non-hermitian fermionic non-interacting models},\ }\bibfield  {journal} {\bibinfo  {journal} {SciPost Physics}\ }\textbf {\bibinfo {volume} {7}},\ \href {https://doi.org/10.21468/SciPostPhys.7.5.069} {10.21468/SciPostPhys.7.5.069} (\bibinfo {year} {2019})\BibitemShut {NoStop}%
\bibitem [{\citenamefont {Eldad}\ \emph {et~al.}(2022)\citenamefont {Eldad}, \citenamefont {Aditya}, \citenamefont {Martin},\ and\ \citenamefont {Susana}}]{Eldad2022Entanglement}%
  \BibitemOpen
  \bibfield  {author} {\bibinfo {author} {\bibfnamefont {B.}~\bibnamefont {Eldad}}, \bibinfo {author} {\bibfnamefont {B.}~\bibnamefont {Aditya}}, \bibinfo {author} {\bibfnamefont {P.}~\bibnamefont {Martin}, \bibfnamefont {B.}},\ and\ \bibinfo {author} {\bibfnamefont {H.}~\bibnamefont {Susana}, \bibfnamefont {F.}},\ }\bibfield  {title} {\bibinfo {title} {Entanglement spectrum in general free fermionic systems},\ }\href {https://doi.org/10.1088/1751-8121/ac5529} {\bibfield  {journal} {\bibinfo  {journal} {J. Phys. A: Math. Theor.}\ }\textbf {\bibinfo {volume} {55}},\ \bibinfo {pages} {135001} (\bibinfo {year} {2022})}\BibitemShut {NoStop}%
\bibitem [{\citenamefont {Assaad}(2015)}]{assaad2015stable}%
  \BibitemOpen
  \bibfield  {author} {\bibinfo {author} {\bibfnamefont {F.~F.}\ \bibnamefont {Assaad}},\ }\bibfield  {title} {\bibinfo {title} {Stable quantum monte carlo simulations for entanglement spectra of interacting fermions},\ }\href@noop {} {\bibfield  {journal} {\bibinfo  {journal} {Physical Review B}\ }\textbf {\bibinfo {volume} {91}},\ \bibinfo {pages} {125146} (\bibinfo {year} {2015})}\BibitemShut {NoStop}%
\bibitem [{\citenamefont {Assaad}\ \emph {et~al.}(2014)\citenamefont {Assaad}, \citenamefont {Lang},\ and\ \citenamefont {Parisen~Toldin}}]{assaad2014entanglement}%
  \BibitemOpen
  \bibfield  {author} {\bibinfo {author} {\bibfnamefont {F.~F.}\ \bibnamefont {Assaad}}, \bibinfo {author} {\bibfnamefont {T.~C.}\ \bibnamefont {Lang}},\ and\ \bibinfo {author} {\bibfnamefont {F.}~\bibnamefont {Parisen~Toldin}},\ }\bibfield  {title} {\bibinfo {title} {Entanglement spectra of interacting fermions in quantum monte carlo simulations},\ }\href@noop {} {\bibfield  {journal} {\bibinfo  {journal} {Physical Review B}\ }\textbf {\bibinfo {volume} {89}},\ \bibinfo {pages} {125121} (\bibinfo {year} {2014})}\BibitemShut {NoStop}%
\bibitem [{\citenamefont {Parisen~Toldin}\ and\ \citenamefont {Assaad}(2018)}]{parisen2018entanglement}%
  \BibitemOpen
  \bibfield  {author} {\bibinfo {author} {\bibfnamefont {F.}~\bibnamefont {Parisen~Toldin}}\ and\ \bibinfo {author} {\bibfnamefont {F.~F.}\ \bibnamefont {Assaad}},\ }\bibfield  {title} {\bibinfo {title} {Entanglement hamiltonian of interacting fermionic models},\ }\href@noop {} {\bibfield  {journal} {\bibinfo  {journal} {Physical review letters}\ }\textbf {\bibinfo {volume} {121}},\ \bibinfo {pages} {200602} (\bibinfo {year} {2018})}\BibitemShut {NoStop}%
\bibitem [{\citenamefont {Cho}\ \emph {et~al.}(2017)\citenamefont {Cho}, \citenamefont {Ludwig},\ and\ \citenamefont {Ryu}}]{Cho2017Universal}%
  \BibitemOpen
  \bibfield  {author} {\bibinfo {author} {\bibfnamefont {G.~Y.}\ \bibnamefont {Cho}}, \bibinfo {author} {\bibfnamefont {A.~W.~W.}\ \bibnamefont {Ludwig}},\ and\ \bibinfo {author} {\bibfnamefont {S.}~\bibnamefont {Ryu}},\ }\bibfield  {title} {\bibinfo {title} {Universal entanglement spectra of gapped one-dimensional field theories},\ }\bibfield  {journal} {\bibinfo  {journal} {Physical Review B}\ }\textbf {\bibinfo {volume} {95}},\ \href {https://doi.org/10.1103/PhysRevB.95.115122} {10.1103/PhysRevB.95.115122} (\bibinfo {year} {2017})\BibitemShut {NoStop}%
\bibitem [{\citenamefont {Mao}\ \emph {et~al.}(2023)\citenamefont {Mao}, \citenamefont {Ding},\ and\ \citenamefont {Zheng}}]{Mao2023Sampling}%
  \BibitemOpen
  \bibfield  {author} {\bibinfo {author} {\bibfnamefont {B.-B.}\ \bibnamefont {Mao}}, \bibinfo {author} {\bibfnamefont {Y.-M.}\ \bibnamefont {Ding}},\ and\ \bibinfo {author} {\bibfnamefont {Y.}~\bibnamefont {Zheng}},\ }\bibfield  {title} {\bibinfo {title} {Sampling reduced density matrix to extract fine levels of entanglement spectrum},\ }\href {https://arxiv.org/abs/2310.16709} {\bibfield  {journal} {\bibinfo  {journal} {arXiv e-prints}\ } (\bibinfo {year} {2023})},\ \Eprint {https://arxiv.org/abs/2310.16709} {arXiv:2310.16709 [cond-mat.str-el]} \BibitemShut {NoStop}%
\bibitem [{\citenamefont {Ingo}(2003)}]{Ingo2003Calculation}%
  \BibitemOpen
  \bibfield  {author} {\bibinfo {author} {\bibfnamefont {P.}~\bibnamefont {Ingo}},\ }\bibfield  {title} {\bibinfo {title} {Calculation of reduced density matrices from correlation functions},\ }\href {https://doi.org/10.1088/0305-4470/36/14/101} {\bibfield  {journal} {\bibinfo  {journal} {J. Phys. A: Math. Gen.}\ }\textbf {\bibinfo {volume} {36}},\ \bibinfo {pages} {L205} (\bibinfo {year} {2003})}\BibitemShut {NoStop}%
\bibitem [{\citenamefont {Peschel}\ and\ \citenamefont {Eisler}(2009)}]{Peschel2009Reduced}%
  \BibitemOpen
  \bibfield  {author} {\bibinfo {author} {\bibfnamefont {I.}~\bibnamefont {Peschel}}\ and\ \bibinfo {author} {\bibfnamefont {V.}~\bibnamefont {Eisler}},\ }\bibfield  {title} {\bibinfo {title} {Reduced density matrices and entanglement entropy in free lattice models},\ }\bibfield  {journal} {\bibinfo  {journal} {Journal of Physics A: Mathematical and Theoretical}\ }\textbf {\bibinfo {volume} {42}},\ \href {https://doi.org/10.1088/1751-8113/42/50/504003} {10.1088/1751-8113/42/50/504003} (\bibinfo {year} {2009})\BibitemShut {NoStop}%
\bibitem [{\citenamefont {D\'Emidio}\ \emph {et~al.}(2024)\citenamefont {D\'Emidio}, \citenamefont {Orús}, \citenamefont {Laflorencie},\ and\ \citenamefont {de~Juan}}]{D2024Universal}%
  \BibitemOpen
  \bibfield  {author} {\bibinfo {author} {\bibfnamefont {J.}~\bibnamefont {D\'Emidio}}, \bibinfo {author} {\bibfnamefont {R.}~\bibnamefont {Orús}}, \bibinfo {author} {\bibfnamefont {N.}~\bibnamefont {Laflorencie}},\ and\ \bibinfo {author} {\bibfnamefont {F.}~\bibnamefont {de~Juan}},\ }\bibfield  {title} {\bibinfo {title} {Universal features of entanglement entropy in the honeycomb hubbard model},\ }\bibfield  {journal} {\bibinfo  {journal} {Physical Review Letters}\ }\textbf {\bibinfo {volume} {132}},\ \href {https://doi.org/10.1103/PhysRevLett.132.076502} {10.1103/PhysRevLett.132.076502} (\bibinfo {year} {2024})\BibitemShut {NoStop}%
\bibitem [{\citenamefont {Wang}\ and\ \citenamefont {Xu}(2023)}]{Wang2023Entanglement}%
  \BibitemOpen
  \bibfield  {author} {\bibinfo {author} {\bibfnamefont {F.-H.}\ \bibnamefont {Wang}}\ and\ \bibinfo {author} {\bibfnamefont {X.~Y.}\ \bibnamefont {Xu}},\ }\bibfield  {title} {\bibinfo {title} {Entanglement r\'enyi negativity of interacting fermions from quantum monte carlo simulations},\ }\href {https://arxiv.org/abs/2312.14155} {\bibfield  {journal} {\bibinfo  {journal} {arXiv e-prints}\ } (\bibinfo {year} {2023})},\ \Eprint {https://arxiv.org/abs/2312.14155} {arXiv:2312.14155 [cond-mat.str-el]} \BibitemShut {NoStop}%
\bibitem [{\citenamefont {Wang}\ \emph {et~al.}(2024{\natexlab{b}})\citenamefont {Wang}, \citenamefont {Wang}, \citenamefont {Ding}, \citenamefont {Mao},\ and\ \citenamefont {Yan}}]{Wang2024Probably}%
  \BibitemOpen
  \bibfield  {author} {\bibinfo {author} {\bibfnamefont {Z.}~\bibnamefont {Wang}}, \bibinfo {author} {\bibfnamefont {Z.}~\bibnamefont {Wang}}, \bibinfo {author} {\bibfnamefont {Y.-M.}\ \bibnamefont {Ding}}, \bibinfo {author} {\bibfnamefont {B.-B.}\ \bibnamefont {Mao}},\ and\ \bibinfo {author} {\bibfnamefont {Z.}~\bibnamefont {Yan}},\ }\bibfield  {title} {\bibinfo {title} {Bipartite reweight-annealing algorithm to extract large-scale data of entanglement entropy and its derivative in high precision},\ }\href {https://arxiv.org/abs/2406.05324} {\  (\bibinfo {year} {2024}{\natexlab{b}})},\ \Eprint {https://arxiv.org/abs/2406.05324} {arXiv:2406.05324 [cond-mat.str-el]} \BibitemShut {NoStop}%
\bibitem [{\citenamefont {Liu}\ and\ \citenamefont {Bryan}(2024)}]{Liu2023Non}%
  \BibitemOpen
  \bibfield  {author} {\bibinfo {author} {\bibfnamefont {Z.}~\bibnamefont {Liu}}\ and\ \bibinfo {author} {\bibfnamefont {K.~C.}\ \bibnamefont {Bryan}},\ }\bibfield  {title} {\bibinfo {title} {Non-equilibrium quantum monte carlo algorithm for stabilizer renyi entropy in spin systems},\ }\href {https://arxiv.org/abs/2405.19577} {\bibfield  {journal} {\bibinfo  {journal} {arXiv e-prints}\ } (\bibinfo {year} {2024})},\ \Eprint {https://arxiv.org/abs/2405.19577} {arXiv:2405.19577 [quant-ph]} \BibitemShut {NoStop}%
\bibitem [{\citenamefont {Wu}\ \emph {et~al.}(2020)\citenamefont {Wu}, \citenamefont {Lu}, \citenamefont {Chung}, \citenamefont {Kao},\ and\ \citenamefont {Grover}}]{Wu2020Entanglement}%
  \BibitemOpen
  \bibfield  {author} {\bibinfo {author} {\bibfnamefont {K.-H.}\ \bibnamefont {Wu}}, \bibinfo {author} {\bibfnamefont {T.-C.}\ \bibnamefont {Lu}}, \bibinfo {author} {\bibfnamefont {C.-M.}\ \bibnamefont {Chung}}, \bibinfo {author} {\bibfnamefont {Y.-J.}\ \bibnamefont {Kao}},\ and\ \bibinfo {author} {\bibfnamefont {T.}~\bibnamefont {Grover}},\ }\bibfield  {title} {\bibinfo {title} {Entanglement renyi negativity across a finite temperature transition: A monte carlo study},\ }\bibfield  {journal} {\bibinfo  {journal} {Physical Review Letters}\ }\textbf {\bibinfo {volume} {125}},\ \href {https://doi.org/10.1103/PhysRevLett.125.140603} {10.1103/PhysRevLett.125.140603} (\bibinfo {year} {2020})\BibitemShut {NoStop}%
\bibitem [{\citenamefont {Liao}(2023)}]{Liao2023Controllable}%
  \BibitemOpen
  \bibfield  {author} {\bibinfo {author} {\bibfnamefont {Y.~D.}\ \bibnamefont {Liao}},\ }\bibfield  {title} {\bibinfo {title} {Controllable incremental algorithm for entanglement entropy in quantum monte carlo simulations},\ }\href {https://arxiv.org/abs/2307.10602} {\bibfield  {journal} {\bibinfo  {journal} {arXiv e-prints}\ } (\bibinfo {year} {2023})},\ \Eprint {https://arxiv.org/abs/2307.10602} {arXiv:2307.10602 [cond-mat.str-el]} \BibitemShut {NoStop}%
\bibitem [{\citenamefont {Wang}\ \emph {et~al.}(2024{\natexlab{c}})\citenamefont {Wang}, \citenamefont {Song}, \citenamefont {Lyu}, \citenamefont {William},\ and\ \citenamefont {Yang}}]{Wang2024Entanglement}%
  \BibitemOpen
  \bibfield  {author} {\bibinfo {author} {\bibfnamefont {T.-T.}\ \bibnamefont {Wang}}, \bibinfo {author} {\bibfnamefont {M.}~\bibnamefont {Song}}, \bibinfo {author} {\bibfnamefont {L.}~\bibnamefont {Lyu}}, \bibinfo {author} {\bibfnamefont {W.-K.}\ \bibnamefont {William}},\ and\ \bibinfo {author} {\bibfnamefont {M.~Z.}\ \bibnamefont {Yang}},\ }\bibfield  {title} {\bibinfo {title} {Entanglement microscopy: Tomography and entanglement measures via quantum monte carlo},\ }\href {https://arxiv.org/abs/2402.14916} {\bibfield  {journal} {\bibinfo  {journal} {arXiv e-prints}\ } (\bibinfo {year} {2024}{\natexlab{c}})},\ \Eprint {https://arxiv.org/abs/2402.14916} {arXiv:2402.14916 [cond-mat.str-el]} \BibitemShut {NoStop}%
\bibitem [{\citenamefont {Ding}\ \emph {et~al.}(2024)\citenamefont {Ding}, \citenamefont {Tang}, \citenamefont {Wang}, \citenamefont {Wang}, \citenamefont {Mao},\ and\ \citenamefont {Yan}}]{ding2024tracking}%
  \BibitemOpen
  \bibfield  {author} {\bibinfo {author} {\bibfnamefont {Y.-M.}\ \bibnamefont {Ding}}, \bibinfo {author} {\bibfnamefont {Y.}~\bibnamefont {Tang}}, \bibinfo {author} {\bibfnamefont {Z.}~\bibnamefont {Wang}}, \bibinfo {author} {\bibfnamefont {Z.}~\bibnamefont {Wang}}, \bibinfo {author} {\bibfnamefont {B.-B.}\ \bibnamefont {Mao}},\ and\ \bibinfo {author} {\bibfnamefont {Z.}~\bibnamefont {Yan}},\ }\bibfield  {title} {\bibinfo {title} {Tracking the variation of entanglement r$\backslash$'enyi negativity: an efficient quantum monte carlo method},\ }\href@noop {} {\bibfield  {journal} {\bibinfo  {journal} {arXiv preprint arXiv:2409.10273}\ } (\bibinfo {year} {2024})}\BibitemShut {NoStop}%
\bibitem [{\citenamefont {Grover}(2013)}]{Grover2013Entanglement}%
  \BibitemOpen
  \bibfield  {author} {\bibinfo {author} {\bibfnamefont {T.}~\bibnamefont {Grover}},\ }\bibfield  {title} {\bibinfo {title} {Entanglement of interacting fermions in quantum monte carlo calculations},\ }\href {https://doi.org/10.1103/PhysRevLett.111.130402} {\bibfield  {journal} {\bibinfo  {journal} {Phys Rev Lett}\ }\textbf {\bibinfo {volume} {111}},\ \bibinfo {pages} {130402} (\bibinfo {year} {2013})}\BibitemShut {NoStop}%
\bibitem [{\citenamefont {Assaad}\ and\ \citenamefont {Evertz}(2008)}]{Assaad2008World-line}%
  \BibitemOpen
  \bibfield  {author} {\bibinfo {author} {\bibfnamefont {F.}~\bibnamefont {Assaad}}\ and\ \bibinfo {author} {\bibfnamefont {H.}~\bibnamefont {Evertz}},\ }\bibinfo {title} {World-line and determinantal quantum monte carlo methods for spins, phonons and electrons},\ in\ \href {https://doi.org/10.1007/978-3-540-74686-7_10} {\emph {\bibinfo {booktitle} {Computational Many-Particle Physics}}},\ \bibinfo {editor} {edited by\ \bibinfo {editor} {\bibfnamefont {H.}~\bibnamefont {Fehske}}, \bibinfo {editor} {\bibfnamefont {R.}~\bibnamefont {Schneider}},\ and\ \bibinfo {editor} {\bibfnamefont {A.}~\bibnamefont {Wei{\ss}e}}}\ (\bibinfo  {publisher} {Springer Berlin Heidelberg},\ \bibinfo {address} {Berlin, Heidelberg},\ \bibinfo {year} {2008})\ pp.\ \bibinfo {pages} {277--356}\BibitemShut {NoStop}%
\bibitem [{\citenamefont {Ye}\ \emph {et~al.}(2016)\citenamefont {Ye}, \citenamefont {Mu},\ and\ \citenamefont {Fan}}]{Ye2016Entanglement}%
  \BibitemOpen
  \bibfield  {author} {\bibinfo {author} {\bibfnamefont {B.-T.}\ \bibnamefont {Ye}}, \bibinfo {author} {\bibfnamefont {L.-Z.}\ \bibnamefont {Mu}},\ and\ \bibinfo {author} {\bibfnamefont {H.}~\bibnamefont {Fan}},\ }\bibfield  {title} {\bibinfo {title} {Entanglement spectrum of su-schrieffer-heeger-hubbard model},\ }\bibfield  {journal} {\bibinfo  {journal} {Physical Review B}\ }\textbf {\bibinfo {volume} {94}},\ \href {https://doi.org/10.1103/PhysRevB.94.165167} {10.1103/PhysRevB.94.165167} (\bibinfo {year} {2016})\BibitemShut {NoStop}%
\bibitem [{\citenamefont {Chandran}\ \emph {et~al.}(2014)\citenamefont {Chandran}, \citenamefont {Khemani},\ and\ \citenamefont {Sondhi}}]{PhysRevLett.113.060501}%
  \BibitemOpen
  \bibfield  {author} {\bibinfo {author} {\bibfnamefont {A.}~\bibnamefont {Chandran}}, \bibinfo {author} {\bibfnamefont {V.}~\bibnamefont {Khemani}},\ and\ \bibinfo {author} {\bibfnamefont {S.~L.}\ \bibnamefont {Sondhi}},\ }\bibfield  {title} {\bibinfo {title} {How universal is the entanglement spectrum?},\ }\href {https://doi.org/10.1103/PhysRevLett.113.060501} {\bibfield  {journal} {\bibinfo  {journal} {Phys. Rev. Lett.}\ }\textbf {\bibinfo {volume} {113}},\ \bibinfo {pages} {060501} (\bibinfo {year} {2014})}\BibitemShut {NoStop}%
\end{thebibliography}%

\end{document}